\documentclass[journal=jacsat,manuscript=article]{achemso}

\setkeys{acs}{keywords = true}
\usepackage[T1]{fontenc}
\usepackage[utf8]{inputenc}
\usepackage{lmodern}
\usepackage{amsmath,amssymb,mathtools}
\usepackage{siunitx}
\usepackage{graphicx}

\usepackage{booktabs}
\usepackage{hyperref}
\usepackage{xcolor}
\usepackage{microtype}
\usepackage[version=3]{mhchem}

\author{Carlos Caro}
\affiliation{Department of Clinical Medicine, Faculty of Health Sciences, UiT-the Arctic University of Norway, 9037 Troms\o, Norway}

\author{Francisco Gámez}
\email{frgamez@ucm.es}
\affiliation{Departamento de Química Física, Fac. Ciencias Químicas, Universidad Complutense de Madrid, E-28040 Madrid, Spain}

\title{Phase-Rotated Altermagnets as Chern Valves for Topological Transport}

\begin{document}

\begin{abstract}
Motivated by the emerging control of Berry-curvature textures in altermagnets, we explore a two-terminal configuration where a topological-insulator film is interfaced with two altermagnetic electrodes whose crystalline phases can be rotated independently. The proximity coupling imprints each altermagnet’s momentum-dependent spin texture onto the Dirac surface states, giving rise to an angular mass whose sign follows the lattice orientation. Adjusting the phase of one electrode redefines this mass pattern, thereby tuning the number and spatial distribution of chiral edge channels. This results in discrete conductance steps and a reversible inversion of the thermoelectric Hall coefficient—achieved without external magnetic fields or net magnetization. A compact Dirac model captures both the quantized switching and its resilience to moderate disorder. Overall, this symmetry-driven mechanism provides a practical and low-dissipation route to programmable topological transport via lattice rotation.
\end{abstract}

Topological insulators (TIs) represent a central platform for investigating quantum phases that support symmetry-protected edge transport and quantized Hall phenomena. \cite{1,2} The first observations of the quantum anomalous Hall effect in magnetically doped TIs firmly established the relationship between symmetry breaking and topological conduction, opening the door to broader families of topological matter, and subsequent developments extended this paradigm to higher-order topological insulators\cite{3,4} as well as to Floquet-engineered and strain-tunable systems.\cite{5,6,7} The ability to manipulate band topology through mechanical or crystalline degrees of freedom now motivates alternative approaches that do not rely on external magnetic fields. A recent and rapidly growing frontier is altermagnetism, a collinear spin order with zero net magnetization but strong momentum-dependent spin splitting enforced by crystalline symmetry.\cite{8,9} Altermagnetic materials such as RuO$_2$, MnTe, and Ca$_3$Ru$_2$O$_7$ display large anomalous Hall effects and strong Berry-curvature multipoles without ferromagnetism.\cite{10,11} These multipoles, rooted in the $C_2$ and $C_4$ symmetry of the lattice, can be reoriented by mechanical strain or shear,\cite{12,13} offering a route to rotate the underlying spin texture \textit{in situ}. Magnetic-proximity experiments on Bi$_2$Se$_3$ interfaces\cite{14} demonstrate that symmetry-controlled exchange fields can open directional Dirac gaps of a few meV, directly linking spin texture and topological transport.
\textcolor{black}{Our proposal is complementary to other recent work on
altermagnet-based spintronics. In particular, de la Barrera and
Núñez have analysed electrical control of the exchange bias
effect at model ferromagnet–altermagnet junctions, where the
altermagnet acts as a collinear antiferromagnetic pinning layer
with spin-split bands and the main observable is the hysteretic
response of the ferromagnet.\cite{PRB_111_174428_2025}
Those proximity geometries focus on tuning the effective
exchange field acting on a ferromagnet by modifying the
altermagnetic order at the interface. In contrast, in the
Chern-valve geometry considered here the altermagnets couple
to the surface Dirac states of a three-dimensional topological
insulator and are used as symmetry-selective sources of an
angular mass. The key control knob is the relative crystalline
phase between two independently rotatable altermagnetic
contacts, which allows us to switch the integer Chern-channel
count at fixed chemical potential and without applying external
magnetic fields. In this sense our mechanism provides a
dynamically reconfigurable, field-free route to topological
transport control that is conceptually distinct from existing
altermagnetic exchange-bias and spin-valve proposals.} Here we combine these ideas into a minimal and experimentally accessible concept: a two-terminal \textit{Chern valve} in which a TI layer is coupled to two altermagnetic electrodes with independently rotatable crystalline phases $\phi_L$ and $\phi_R$. The operating principle is illustrated in Fig.~1(a): a TI strip contacted by two altermagnets (AMs) transfers, via spin–orbit proximity, their symmetry-dependent spin textures to the Dirac surface states, generating an angular mass $m(\theta;\phi)$ whose sign alternates with the local lattice orientation. A finite phase offset $\Delta\varphi=\phi_R-\phi_L$ creates regions where $m_L(\theta)m_R(\theta)<0$, fulfilling the Jackiw–Rebbi criterion\cite{15,16} and hosting chiral one-dimensional channels at the interfaces.

The essential physics is captured by a minimal Dirac model describing a two-dimensional TI surface proximized by two altermagnets with independently rotated crystalline phases:
\begin{equation}
H_{\mathrm{TI}}(\mathbf{k})=\hbar v_F(k_x\sigma_y-k_y\sigma_x)+m(\theta;\phi)\,\sigma_z,
\label{eq:H}
\end{equation}
where $\mathbf{k}=(k_x,k_y)$ is the crystal momentum, $\theta=\mathrm{atan2}(k_y,k_x)$ the azimuthal angle, $v_F$ the Fermi velocity, and $\sigma_{x,y,z}$ are Pauli matrices. Because typical altermagnets exhibit $C_2$ or $C_4$ spin-rotation symmetries, their Berry-curvature multipoles map onto the harmonic angular dependence of the mass term:
\begin{equation}
m(\theta;\phi)=m_0+m_2\cos[2(\theta-\phi)]+m_4\cos[4(\theta-\phi)],
\end{equation}
where $m_0$, $m_2$, and $m_4$ are real amplitudes proportional to the interfacial exchange and spin–orbit coupling strength. The $m_2$ and $m_4$ coefficients encode the dipolar ($C_2$) and quadrupolar ($C_4$) Berry-curvature multipoles of the altermagnet. In the $\mathbf{d}$-vector notation with $\mathbf{d}(\mathbf{k};\phi)=(-\hbar v_F k_y,\ \hbar v_F k_x,\ m(\theta;\phi))$, the band-resolved Berry curvature is:
\begin{equation}
\Omega_{\pm,z}(\mathbf{k};\theta)=\mp\frac{1}{2}\,\frac{\mathbf{d}\cdot\bigl(\partial_{k_x}\mathbf{d}\times\partial_{k_y}\mathbf{d}\bigr)}{|\mathbf{d}|^3}.
\end{equation}
The intrinsic anomalous Hall conductivity follows from integrating the Berry curvature over the Brillouin zone:
\begin{equation}
\sigma_{xy}=-\frac{e^2}{\hbar}\sum_{n=\pm}\int_{\mathrm{BZ}}\frac{d^2\mathbf{k}}{(2\pi)^2}\,f(\varepsilon_{n\mathbf{k}},\mu,T)\,\Omega_{n,z}(\mathbf{k}),
\end{equation}
where $\varepsilon_{n\mathbf{k}}$ is the energy dispersion and $f(\varepsilon_{n\mathbf{k}},\mu,T)$ is the Fermi–Dirac distribution. The thermoelectric Hall coefficient satisfies the low-temperature Mott relation:
\begin{equation}
\alpha_{xy}=-\frac{\pi^2k_B^2T}{3e}\left(\frac{\partial\sigma_{xy}}{\partial\mu}\right)_{T\to 0},
\end{equation}
which holds for $T\lesssim 10$ K in typical AM|TI stacks. Because $m(\theta;\phi)$ rotates rigidly with the crystalline phase, both $\sigma_{xy}$ and $\alpha_{xy}$ inherit its $C_2/C_4$ symmetry, providing a direct symmetry-protected electrical and thermoelectric readout of the Chern-active regions.

\begin{figure*}[t]
\centering
\includegraphics[width=0.95\linewidth]{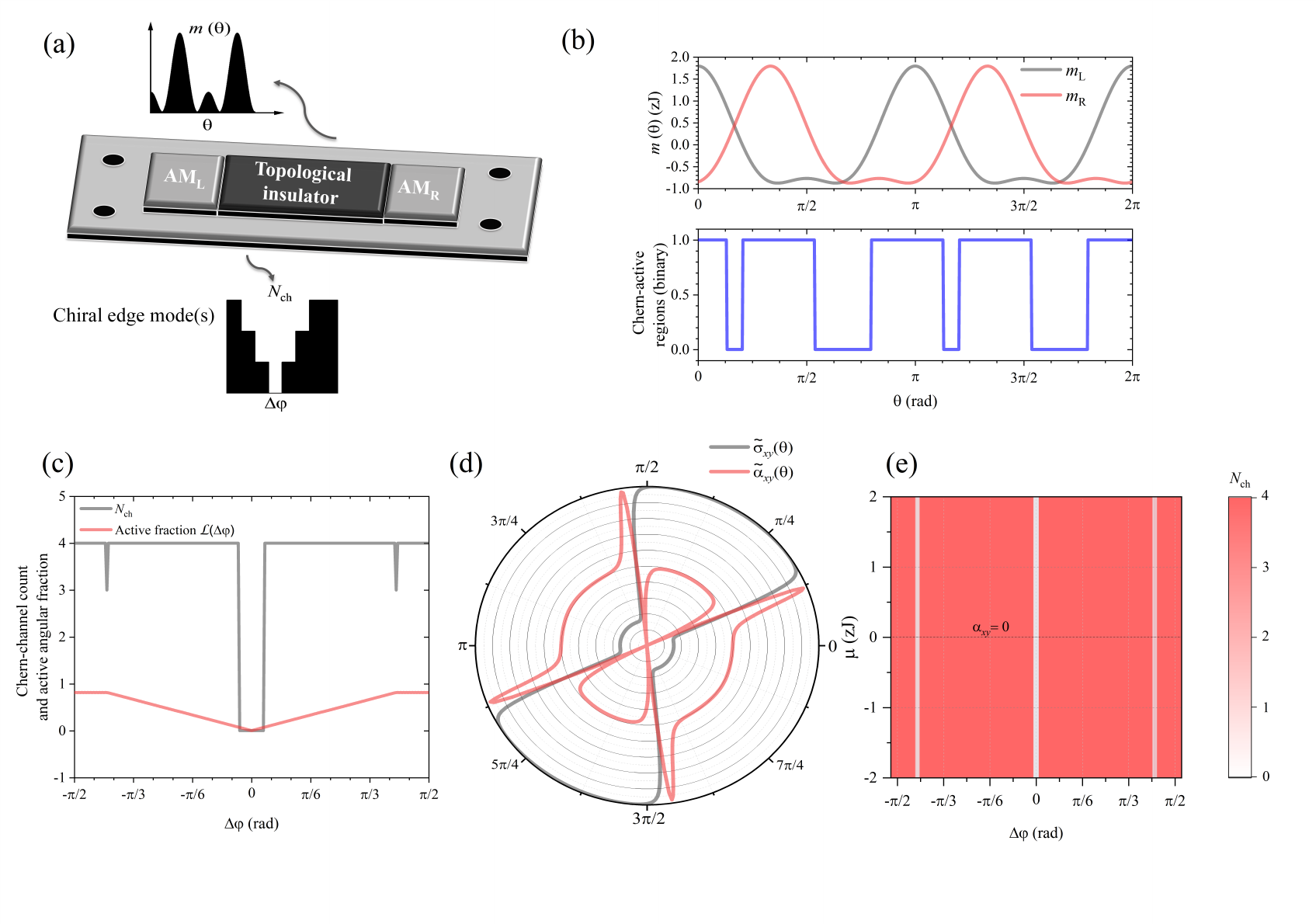}
\caption{{\color{black}\textbf{Chern-valve concept and angular-mass topology.}
(a) Schematic of the AM$|$TI$|$AM junction with altermagnets at
crystalline phases $\varphi_L$ and $\varphi_R$ inducing angle-dependent
masses $m_{L/R}(\theta)$ on the Dirac surface states. A sign inversion
across an interface ($m_L m_R<0$) creates a chiral edge mode (red arrow).
(b) Upper panel: angular masses $m_L(\theta)$ and $m_R(\theta)$ for
$\Delta\varphi=\varphi_R-\varphi_L=\pi/3$, showing two principal lobes
reflecting the $C_2$ and $C_4$ harmonics. Lower panel: binary signal
marking Chern-active sectors where $m_L(\theta)m_R(\theta)<0$.
(c) Integer Chern-channel count $N_{\mathrm{ch}}$ (gray, step-like) and
active angular fraction $\mathcal{L}(\Delta\varphi)$ (red, continuous) as
a function of phase difference $\Delta\varphi$, showing quantized plateaus.
The fraction $\mathcal{L}$ is normalised such that
$\mathcal{L}=1$ corresponds to full $2\pi$ angular coverage.
Numerical simulations employ representative parameters for
RuO$_2$|Bi$_2$Se$_3$ interfaces:
$m_2\approx 2~\mathrm{meV}$, $m_4\approx 0.5~\mathrm{meV}$,
Fermi velocity $v_F=5\times10^{5}~\mathrm{m/s}$, and
temperature $T=15~\mathrm{K}$.
(d) Normalised anomalous Hall and thermoelectric Hall conductivities
$\tilde\sigma_{xy}(\varphi)$ and $\tilde\alpha_{xy}(\varphi)$ exhibiting
$C_2$/$C_4$ harmonic phase shifts locked to $\varphi$, with
$\tilde\sigma_{xy}(\varphi)=\sigma_{xy}(\varphi)/
\max_{\varphi}|\sigma_{xy}|$ and
$\tilde\alpha_{xy}(\varphi)=\alpha_{xy}(\varphi)/
\max_{\varphi}|\alpha_{xy}|$.
(e) Phase diagram of the Chern-valve response
$N_{\mathrm{ch}}(\mu,\Delta\varphi)$ across the $(\mu,\Delta\varphi)$
plane. White regions correspond to $N_{\mathrm{ch}}=0$ (aligned
altermagnetic masses without sign inversion), while red regions denote
finite $N_{\mathrm{ch}}$ where chiral edge channels are active; the colour
scale encodes the integer channel count. The purple contour marks the
condition $\alpha^{\mathrm{tot}}_{xy}=0$, separating regions of opposite
transverse thermoelectric polarity. This map demonstrates that the
quantised channel topology and the thermo-Hall polarity remain robust
over a wide range of electrochemical conditions.}}
\label{fig:1}
\end{figure*}

The quantized topological channels emerge when the angular mass changes sign between the two altermagnetic contacts. Figure~1(b) illustrates how the relative crystalline phase $\Delta\varphi$ controls chiral-sector formation in the TI channel. The upper panel shows the angular masses $m_L(\theta)$ and $m_R(\theta)$, each displaying two principal positive and negative lobes within a $2\pi$ cycle, reflecting the superposition of $C_2$ and $C_4$ harmonics. With the red curve ($m_R$) offset by $\Delta\varphi=\pi/3$ from the gray one ($m_L$), the zeros alternate along $\theta$, creating four regions where $m_L(\theta)$ and $m_R(\theta)$ have opposite signs. The lower panel marks the Chern-active sectors where $m_L(\theta)m_R(\theta)<0$. For each $\theta$, this binary signal identifies sectors satisfying the Jackiw–Rebbi condition and hosting one-dimensional chiral edge states.

The number of active sectors is obtained by counting the sign reversals:
\begin{equation}
N_{\mathrm{ch}}=\mathrm{nint}\left[\frac{1}{2\pi}\int_0^{2\pi}d\theta\,\Theta[-m_L(\theta)m_R(\theta)]\right],
\end{equation}
with the continuous active angular fraction:
\begin{equation}
\mathcal{L}(\Delta\varphi)=\frac{1}{2\pi}\int_0^{2\pi}d\theta\,\Theta[-m_L(\theta)m_R(\theta)],
\end{equation}
\textcolor{black}{where $\Theta(x)$ denotes the Heaviside
step function, with $\Theta(x>0)=1$ and $\Theta(x\le 0)=0$, and that
$\mathrm{nint}[x]$ returns the nearest integer to $x$. In practice we evaluate
$N_{\mathrm{ch}}$ by discretising $\theta$ into $N_\theta$ points
and counting the connected angular sectors where
$m_L(\theta)m_R(\theta)<0$ with a minimal width $w_{\min}$, which
is equivalent to Eq.~(6) in the continuum limit.} For $\Delta\varphi=\pi/3$, four ``on'' intervals appear, giving $N_{\mathrm{ch}}=4$ and $\mathcal{L}\approx 0.5$, meaning roughly half the Fermi contour contributes to topological transport.

Figure~1(c) shows the phase-offset dependence. The gray trace (integer $N_{\mathrm{ch}}$) remains at four for most offsets, collapsing to zero near $\Delta\varphi=0$ where the masses are aligned and no sign inversion occurs. Small notches where $N_{\mathrm{ch}}=3$ appear near $\Delta\varphi=\pm\pi/3$, when two sign-changing boundaries merge into a tangential zero of $m_L(\theta)m_R(\theta)$, momentarily suppressing one active sector. The red curve ($\mathcal{L}$) varies smoothly, reaching its minimum at alignment. Both observables reveal the same mechanism: as the relative crystalline phase increases, additional opposite-sign sectors emerge sequentially, enabling stepwise tuning of the topological channels.

Figure~1(d) displays the angular dependence of the intrinsic Hall responses. Both the anomalous Hall and thermoelectric Hall coefficients (normalized to emphasize relative phase and symmetry) follow the fourfold $C_2/C_4$ pattern as the mass term but shift in phase with $\phi$, providing a direct symmetry-locked electrical and thermoelectric signature and confirming their common Berry-curvature origin through the coincidence of peaks with angular sectors where $m_Lm_R<0$.

The robustness of the topological quantization across the full accessible range of chemical potentials and phase offsets is captured in Fig.~1(e), which displays a phase diagram mapping $N_{\mathrm{ch}}(\mu,\Delta\varphi)$ across the $(\mu,\Delta\varphi)$ plane. The diagram reveals quantized plateaus as a function of both the Fermi level (chemical potential $\mu$) and the relative crystalline phase. The white regions correspond to $N_{\mathrm{ch}}=0$, where the left and right altermagnetic masses are aligned and no sign inversion occurs, whereas the red sectors indicate finite $N_{\mathrm{ch}}$ representing active chiral-valve configurations. The purple contour marks the condition $\alpha_{xy}^{\mathrm{tot}}=0$, separating regions of opposite transverse thermoelectric polarity. This map demonstrates that the quantized channel topology and its associated thermo-Hall response remain robust over a wide range of electrochemical conditions, which is crucial for device operation.

The proposed geometry can be implemented using established thin-film growth and strain-control techniques. High-quality topological-insulator Bi$_2$Se$_3$ films (50--100 nm) can be grown by molecular beam epitaxy on various substrates, with excellent structural quality demonstrated through rocking-curve linewidths below 15 arcsec and clear layer thickness fringes.\cite{20,21} For the altermagnetic contact layers, RuO$_2$ (10--20 nm) and $\alpha$-Fe$_2$O$_3$ (hematite) films are deposited by pulsed-laser deposition on compatible oxide substrates. The lattice matching between Bi$_2$Se$_3$ and typical substrates (SrTiO$_3$, InP, Al$_2$O$_3$) exhibits lattice mismatch below 3\%, supporting coherent heteroepitaxial growth via strain-mediated van der Waals interactions.\cite{22,23} Independent in-plane strain control can be achieved through piezoelectric or flexible substrates and actuators. Experimental observations confirm that controlled strain of $\approx 1\%$ is sufficient to induce phase-dependent changes in collinear magnetic order in both RuO$_2$ and MnTe, with such strain amplitudes being reversible and non-hysteretic across multiple cycles.\cite{17,18,19} This level of strain is achievable through standard strain-engineering techniques in oxide heterostructures and constitutes an experimentally realistic switching mechanism for the proposed Chern valve.
\textcolor{black}{To quantify the required crystalline rotation and strain, we have evaluated
the channel count $N_{\mathrm{ch}}(\Delta\varphi)$ as a function of the
relative phase between the two altermagnets,
$\Delta\varphi = \varphi_R - \varphi_L$, using the same representative
parameters as in Fig.~\ref{fig:1}. Imposing a minimal angular width of
$w_{\min} = 6^\circ$ for a conducting sector, we find that the Chern valve
remains in a fully blocked regime ($N_{\mathrm{ch}} = 0$) for
$|\Delta\varphi| \lesssim 6^\circ$, while a four–channel state with
$N_{\mathrm{ch}} = 4$ emerges for $|\Delta\varphi| \gtrsim 6.5^\circ$.
In other words, a relative misalignment of the Néel vectors by only
$\sim 5$–$10^\circ$ is sufficient to switch between the ``off'' and ``on''
topological plateaus. Recent experimental and theoretical works on epitaxial
RuO$_2$ and MnTe indicate that uniaxial strains of order
$\varepsilon \sim 1\,\%$ can already drive a repopulation of domains and a
rotation of the altermagnetic spin texture by angles of the order of a few
tens of degrees.\cite{17,18,19} Within this range, a full phase difference
$\Delta\varphi \simeq \pi/2$, sufficient to traverse one conductance plateau,
would correspond to strains of order $\varepsilon \sim 2$–$3\,\%$, still well
within the elastic window of oxide and chalcogenide thin films on
piezoelectric substrates. Additionally, using DFT-based Wannier tight-binding parameters for RuO$_2$~\cite{10}
and experimental estimates of proximity-induced exchange gaps in
magnetically gapped Bi$_2$Se$_3$ surface states~\cite{14}, we can obtain
a simple order-of-magnitude estimate for the angular mass in a
RuO$_2$|Bi$_2$Se$_3$ heterostructure. Taking a representative interfacial
exchange splitting $\Delta_{\mathrm{ex}}\approx 5$–$10~\mathrm{meV}$,
a $\mathbf{k}\cdot\mathbf{p}$ downfolding of the bulk altermagnetic
spin texture onto the TI surface yields harmonic amplitudes of order $m_2\simeq 0.6\,\Delta_{\mathrm{ex}}$
and $m_4\simeq 0.3\,\Delta_{\mathrm{ex}}$. For the above range of
$\Delta_{\mathrm{ex}}$ this gives $m_2\approx 3$–$6~\mathrm{meV}$ and
$\,m_4\approx 1.5$–$3~\mathrm{meV}$, i.e.\ meV-scale masses fully
consistent with the parameters used in our simulations and compatible
with the Dirac-gap sizes reported in Refs.~\cite{14,20,21}. This
confirms that the angular masses required for Chern-valve operation are
achievable, for instance, in realistic RuO$_2$|Bi$_2$Se$_3$ devices. We emphasise that the strain amplitudes considered here,
$\varepsilon \sim 1$–$2\,\%$, correspond to in-plane epitaxial or
piezoelectric strain used to reorient the altermagnetic crystalline
phase, rather than to drive a bulk topological transition in the TI
itself. In our description the three-dimensional TI remains in the same
$Z_2$ topological phase throughout; strain only enters via the
altermagnets, by modifying the orientation and magnitude of the
proximity-induced angular mass $m(\theta;\varphi)$ at the surface.
The bulk Dirac gap and Fermi velocity of the TI are kept fixed and well
within the topological regime, so that the Chern-valve operation is
entirely controlled by boundary conditions. Much stronger or
non-uniform deformations, such as large out-of-plane strain capable of
closing and reopening the bulk gap of the TI, could in principle trigger
a separate topological phase transition, but such regimes lie outside
the operating window of the present proposal.}

In the phase-rotated AM$\mid$TI$\mid$AM junction, transport within the interfacial gap is
carried by one-dimensional chiral channels that appear whenever the angular masses have
opposite sign, $m_L(\theta)m_R(\theta)<0$. For fixed $(\mu,\Delta\varphi)$,
let $N_{\mathrm{ch}}(\mu,\Delta\varphi)$ be the number of such channels. The two-terminal
conductance in the Landauer picture reads
\begin{equation}
G(\mu,\Delta\varphi,T)
=\frac{e^2}{h}\sum_{i=1}^{N_{\mathrm{ch}}(\mu,\Delta\varphi)}\mathcal{T}_i(\mu,T),
\label{eq:Landauer_G}
\end{equation}
where $\mathcal{T}_i$ is the transmission of channel $i$.
In the clean, low-temperature limit and for well-matched contacts we have
$\mathcal{T}_i\!\to\!1$, hence
\begin{equation}
G(\mu,\Delta\varphi,0)\;=\;\frac{e^2}{h}\,N_{\mathrm{ch}}(\mu,\Delta\varphi).
\label{eq:quantized_G}
\end{equation}
As the relative crystalline phase $\Delta\varphi$ is tuned, the set of angles $\theta$ that
satisfy $m_L(\theta)m_R(\theta)<0$ changes discretely: each creation/annihilation of a
Chern-active sector adds/removes one chiral mode, producing conductance steps of height
$\Delta G=e^2/h$. No additional factor of two appears because each chiral channel is singly
degenerate (the exchange-induced gap on the TI surface breaks Kramers degeneracy and lifts
spin doubling). At finite temperature or with moderate disorder, $\mathcal{T}_i<1$ and the
plateaus acquire a slight slope, but their spacing in units of $e^2/h$ remains set by
$N_{\mathrm{ch}}$. Consequently, two-terminal conductance measurements should reveal discrete plateaus separated by
$e^2/h$ as the relative crystalline phase is tuned through strain, in accordance with
Eqs.~\eqref{eq:Landauer_G}–\eqref{eq:quantized_G}. Simultaneous thermoelectric characterization using standard micro-heater and thermometer geometries can detect the predicted sign inversion of $\alpha_{xy}$.
The absence of net magnetization eliminates parasitic Hall offsets and stray-field effects,
allowing the geometric nature of the switching to be isolated unambiguously.
The channel-switching energy scale of order $1~\mathrm{meV}$ implies operational temperatures
up to $\sim 15~\mathrm{K}$ (consistent with $k_{\mathrm{B}}T \lesssim 1~\mathrm{meV}$),
accessible with standard cryogenic setups.
Reversible piezoelectric actuation enables dynamic tuning rates in the microsecond range,
making the device suitable for low-power, symmetry-controlled topological logic elements.
\textcolor{black}{Although we have illustrated the Chern-valve mechanism using
$C_2$/$C_4$ altermagnets, the construction is not restricted to these
symmetries. For a $C_6$ altermagnet the angular mass would acquire an
additional harmonic of the form
$m_6\cos[6(\theta-\varphi)]$, leading to six positive and six negative
lobes of $m(\theta;\varphi)$ around the Fermi contour. A finite phase
offset $\Delta\varphi$ between two such contacts still produces angular
sectors where $m_L(\theta)m_R(\theta)<0$ and thus hosts chiral interface
channels; only the number and angular width of the Chern-active sectors
change compared to the $C_2$/$C_4$ case. A quantitative analysis of
$N_{\mathrm{ch}}(\Delta\varphi)$ for concrete $C_6$ altermagnets is left
for future work, but the symmetry considerations that underlie the
Chern-valve mechanism apply equally to recently identified hexagonal
altermagnets}
\textcolor{black}{Several practical limitations merit explicit discussion. First, our minimal
Dirac model assumes sharp interfaces and an angular mass containing only the
leading $C_2$/$C_4$ harmonics. In this description the robustness of the
Chern-valve plateaus is controlled primarily by the sign structure of
$m(\theta)$ and by the presence of an interfacial gap, rather than by a
fine tuning of individual harmonic amplitudes. In a real device,
symmetry-breaking distortions, including higher-order harmonics
($m_6$, $m_8$, $\dots$), interface roughness, interdiffusion and intermixing
at the AM$\mid$TI boundaries will inevitably introduce scattering and distort
the Berry-curvature texture. Moderate distortions of this kind deform the
angular regions where $m_L(\theta)m_R(\theta)<0$, broadening the switching
transitions in the phase diagram of Fig.~1(e) and reducing the transmission
of individual channels ($0<\mathcal{T}_i<1$), so that the plateaus become
slightly rounded while their spacing in units of $e^2/h$ remains fixed by the
integer channel count $N_{\mathrm{ch}}$. A fully microscopic treatment of
strong disorder in the AM and TI regions, including explicit band-structure
or scattering-matrix calculations, lies beyond the present scope but would be
highly valuable for device-level optimisation.}

\textcolor{black}{Second, the predicted switching energy scale ($\approx 1$~meV) sets the
operational temperature limit to $T \lesssim 15$~K, constraining practical
device deployment to cryogenic platforms. This limitation is shared with
other geometric topological switches and is still less stringent than for
many superconducting approaches. Third, achieving independent strain control
across both altermagnetic contacts demands either separate piezoactuators
(increasing complexity and power) or spatially resolved strain patterning
via lithography, both of which are technologically feasible but will require
further optimisation. Fourth, the model neglects lattice imperfections, point
defects and thermal magnon excitations in the altermagnet, which can
renormalise the exchange coupling and partly mask the predicted quantisation
under ambient conditions. Fifth, the Fermi-level tunability shown in
Fig.~1(e) assumes clean charge-carrier accumulation; in real devices,
band-bending, trap states and back-gate leakage will distort the
$\mu(\Delta\varphi)$ map and reduce the visibility of individual plateaus.
Despite these challenges, the underlying protection mechanism, rooted in
Jackiw–Rebbi zero-mode formation and in the topological sign structure of
the mass, is robust to small perturbations that preserve the primary
$C_2$ or $C_4$ symmetry and keep the interfacial gap open. Experimental
demonstration will be crucial to determine the actual quantisation tolerance
margins and to refine material choices and growth protocols}

\begin{table}[t]
\centering
\caption{Comparison of topological-switching mechanisms and their experimental benchmarks.}
\label{tab:comparison}
\begin{tabular}{lccc}
\toprule
Feature & Magnetic field\cite{1} & Floquet drive\cite{5,6} & This work (strain/rotation) \\
\midrule
Dissipation & High (Oersted, eddy) & Moderate–high (optical) & Minimal (geometric) \\
Speed & ms range & fs–ps range & $\mu$s range (piezo) \\
Reversibility & Often hysteretic & Periodic & Fully reversible \\
Operation temperature & 1–300 K & Variable & $<15$ K \\
Magnetization & Yes & No & No \\
Physical handle & Magnetic field & Light & Lattice rotation (strain) \\
\bottomrule
\end{tabular}
\end{table}

The Chern valve represents a distinct switching paradigm compared to established mechanisms. Magnetic-field-controlled Chern insulators demand strong external fields and typically exhibit hysteretic behavior,\cite{1} while Floquet topological engineering requires high-frequency optical modulation with inherent dissipation.\cite{5,6} This approach achieves fully reversible control with minimal dissipation via static lattice rotation. Table~1 compares operational characteristics and literature benchmarks.\cite{1,5,6,24,25} Extensions of this concept stablish a route toward strain-tunable topological logic and spin-orbit device platforms

\appendix
\section{Numerical Grid and Convergence Details}

All numerical simulations were performed on dense, symmetry-adapted grids to ensure full convergence of both the topological and thermoelectric quantities. The momentum-space integrals were computed using
$N_k = 300$, $N_\theta = 601$, and $N_{\Delta\phi} = 241$, uniformly sampling the Brillouin zone and the interlayer phase difference. The chemical potential was discretized into $N_\mu = 151$ covering either the gapped window
\([-6,6]~\mathrm{meV}\)
or the extended range
\([-15,15]~\mathrm{meV}\)
used in the wide-band phase diagrams. The angular integration over the scattering plane was limited to a minimal width of $w_{\mathrm{min}} = 6^{\circ}$, which ensures stability of the polar plots. Convergence tests were carried out by doubling the sampling densities
\((N_k, N_\theta)\),
leading to variations below
\(10^{-3}\)
in all integrated quantities. All figures presented in this work correspond to these numerical parameters unless explicitly stated otherwise.

\section*{Author Information}
\noindent\textbf{Corresponding Author.} E-mail: frgamez@ucm.es

\section*{Acknowledgment}
This project has been funded by grants PID2022-136919NA-C33 of the Ministry of Science, Innovation and Universities MCIN/AEI/ 10.13039/501100011033.


\begin{thebibliography}{99}

\bibitem{1} Chang, C.-Z.; et al. Experimental Observation of the Quantum Anomalous Hall Effect in a Magnetic Topological Insulator. \textit{Science} \textbf{2013}, 340, 167–170.

\bibitem{2} Qi, X.-L.; Zhang, S.-C. Topological Insulators and Superconductors. \textit{Rev. Mod. Phys.} \textbf{2011}, 83, 1057.


\bibitem{3} Peterson, C. W.; Benalcazar, W. A.; Hughes, T. L.; Bahl, G. A Quantized Microwave Quadrupole Insulator with Topologically protected Corner States. \textit{Nature} \textbf{2018}, 555, 346–350.


\bibitem{4} Ezawa, M. Topological Switch between Second-Order Topological and Trivial Insulators. \textit{Phys. Rev. Lett.} \textbf{2018}, 120, 026801.

\bibitem{5} Rechtsman, M. C.; et al. Photonic Floquet Topological Insulators. \textit{Nature} \textbf{2013}, 496, 196–200.

\bibitem{6} McIver, J. W.; et al. Light-Induced Anomalous Hall Effect in Graphene. \textit{Nat. Phys.} \textbf{2020}, 16, 38–41.

\bibitem{7} Peotta, S.; Törmä, P. Superfluidity in Topologically Nontrivial Flat Bands. \textit{Nat. Commun.} \textbf{2015}, 6, 8944.

\bibitem{8} Šmejkal, L.; Sinova, J.; Jungwirth,T. Beyond Conventional Ferromagnetism and Antiferromagnetism: A Phase with Nonrelativistic Spin and Crystal Rotation Symmetry. \textit{Phys. Rev. X} \textbf{2022}, 12, 031042.

\bibitem{9} Šmejkal, L.; et al. Emerging Research Landscape of Altermagnetism. \textit{Phys. Rev. X} \textbf{2022}, 12, 040501.

\bibitem{10} Guo, Y.; et al. Direct and Inverse Spin Splitting Effects in Altermagnetic RuO$_2$. \textit{Adv. Sc.} \textbf{2024}, 11(25), 2400967.

\bibitem{11}Cheng-Ping, Z. et al. Higher-order nonlinear anomalous Hall effects induced by Berry curvature multipoles. \textit{Phys. Rev. B} \textbf{2023}, 107, 115142.

\bibitem{12} Fu, Z.; et al. Multiple Topological Phases Controlled via Strain in Two-Dimensional Altermagnets. DOI: 10.48550/arXiv.2507.22474 (2025)

\bibitem{13} Cenker, J.; et al. Reversible strain-induced magnetic phase transition in a van der Waals magnet. \textit{Nat. Nanotechnol} \textbf{2022}, 17, 256-261.


\bibitem{PRB_111_174428_2025} De la Barrera, G.; Nunez, A. S. Electrical control of the exchange bias effect at model ferromagnet-altermagnet junctions. \textit{Phys. Rev. B.} \textbf{2025}, 111, 174428


\bibitem{14} Mogi, M.; et al. Current-induced Switching of Proximity-Induced Ferromagnetic Surface States in a Topological Insulator. \textit{Nat. Comm} \textbf{2021}, 1404.



\bibitem{15} Jackiw, R.; Rebbi, C. Solitons with Fermion Number 1/2. \textit{Phys. Rev. D} \textbf{1976}, 13, 3398–3409.

\bibitem{16} Semenoff, G. W. Condensed-Matter Simulation of a Three-Dimensional Anomaly. \textit{Phys. Rev. Lett.} \textbf{1984}, 53, 2449–2452.

\bibitem{17} Wickramaratne, D.; et al. Effects of altermagnetic order, strain and doping in RuO$_2$.  \textit{J. Mat. Chem. C} \textbf{2026}. DOI: 10.1039/D5TC02898A.

\bibitem{18} Smolenski, G.; et al. Strain-tunability of the multipolar Berry curvature in layered altermagnets. DOI:10.48550/arXiv.2509.21481 (2025).

\bibitem{19} Takahashi, T.; et al. Elasto-Hall conductivity and the anomalous Hall effect in strained altermagnetic materials. \textit{Phys. Rev. B} \textbf{2025}, 111, 184408.

\bibitem{20} Schreyeck, S.; et al. Molecular beam epitaxy of high structural quality Bi$_2$Se$_3$ on lattice matched InP(111) substrates. \textit{Phys. Rev. B} \textbf{2013}, 87, 165139.

\bibitem{21} Jerng, S. K.; et al. Ordered Growth of Topological Insulator Bi$_2$Se$_3$ Thin Films on Amorphous SiO$_2$ by Molecular Beam Epitaxy. \textit{J. Phys. Chem. C} \textbf{2013}, 117, 5043–5047.

\bibitem{22} Xu, S.; et al. van der Waals Epitaxial Growth of Atomically Thin Bi$_2$Se$_3$ and Thickness-Dependent Topological Phase Transition. \textit{Nano Lett.} \textbf{2015}, 15(4), 2645–2651


\bibitem{23} Wang, Z.; Law, S. Optimization of the growth of the van der Waals materials Bi$_2$Se$_3$ and (Bi$_0.5$In$_0.5$)$_2$Se$_3$ by molecular beam epitaxy. arXiv:2107.09771 (2021).

\bibitem{24} Deng, Y.; et al. Quantum anomalous Hall effect in intrinsic magnetic topological insulator MnBi$_2$Te$_4$. \textit{Science} \textbf{2020}, 367, 895–900.

\bibitem{25} Serlin, M.; et al. Intrinsic quantized anomalous Hall effect in a moiré heterostructure. \textit{Science} \textbf{2020}, 367, 900–903.

\end{thebibliography}
\end{document}